\documentclass[aps,prl,superscriptaddress,showpacs,twocolumn]{revtex4}
\usepackage{amsfonts}
\usepackage{graphicx}
\usepackage{graphicx}
\usepackage{amsmath}
\usepackage{amssymb}

\begin{document}

\title{Consistency in Formulation of Spin Current and Torque Associated with
a Variance of Angular Momentum}
\author{Yong Wang}
\affiliation{Institute of Physics, Chinese Academy of Sciences, Beijing 100080, China}
\affiliation{School of Physics, Peking University, Beijing 100871, China}
\author{Ke Xia}
\affiliation{Institute of Physics, Chinese Academy of Sciences, Beijing 100080, China}
\author{Zhao-Bin Su}
\affiliation{Institute of Theoretical Physics, Chinese Academy of Sciences, Beijing
100080, China}
\author{Zhongshui Ma$^a$}
\affiliation{School of Physics, Peking University, Beijing 100871, China}
\begin{abstract}
Stimulated generally by recent interest in the novel spin Hall
effect, the nonrelativistic quantum mechanical conserved currents,
taken into account of spin-orbit coupling, are rigorously
formulated based on the symmetries of system and Noether' theorem.
The quantum mechanical force on the spin as well as the torque
associated with the variance of angular momentum are obtained.
Consequently, the kinetic interpretation of the variances of spin
and orbit angular momentum currents implies a torque on the
"electric dipole" associated with the moving spin. The bearing of
the force and the torque on the properties of spin current in a
two-dimensional electron gas with the Rashba spin-orbit
interaction is discussed.
\end{abstract}

 \maketitle

The theoretical prediction of spin Hall effect
(SHE)\cite{Hirsch99,Murakami03,Sinova04,Shen04} has drawn
considerable interest recently. The spin Hall current comes
forward from the assembly effect that the carriers with opposite
spins move in the opposite directions perpendicular to the
electric field. Equivalently, a spin imbalance between its
opposite flows is electrically generated via the spin-orbit (SO)
coupling in the confined semiconductor. The manipulation of
transverse force on the moving carriers with different spin
orientations, i.e., pushes spin-up and -down carriers in the
opposite directions, is understood as a dynamical issue of
spintronics\cite{Ryu96,Nikolic05,Shen05}. By using the optical
methods, the experimental detections of spin accumulation in
connection with SHE have recently been
reported\cite{Kato04,Wunderlich05}. The effect of impurity
scattering on SHE has also been
investigated\cite{Inoue04,Mishchenko04,Sheng05}.

However, a more fundamental question is emerged in the
conventional understanding of SHE. The validity of the description
for various physical properties of SHE depends on the correctness
of definition for the spin current
operator\cite{Sun05,Zhang05,Wang05}. Since the translational and
rotational motions are associated to the force and torque, without
a rigorous definition of spin current, in some sense, the
discussion about the convenient power source for the spin current
is also somewhat arbitrary. Currently, the spin current operator
is practically defined in analogy with that of charge current as a
tensor operator of symmetric multiplying of Pauli matrix and
velocity, $\mathbb{J} _{\alpha }^{\mathcal{S}}=(\hbar/2)\{\sigma
_{\alpha },\mathbf{v}\}$\cite{Murakami03,Sinova04,Shen04}.
Nevertheless it does not satisfy a continuity
equation\cite{Sun05,Zhang05,Wang05}, ${
\partial {\rho }_{\alpha }^{
\mathcal{S}}}/{\partial t}+\mathbf{\nabla }\cdot {\mathbb{J}_{\alpha }^{%
\mathcal{S}}}\neq 0$, where ${\rho }_{\alpha
}^{\mathcal{S}}=(\hbar /2)\psi ^{\dag }\sigma _{\alpha }\psi$ is a
spin density. Unlike the charge current, the conserved spin
current can not be obtained within the context of Schr\"{o}dinger
equation with the addition of the SO interaction.

Actually, spin motion is intimately related to the orbital angular
momentum of spin carriers. If we identify spin as the internal
angular momentum of carriers, the orbit angular momentum can then
be regarded as a dynamical property of carrier motion. The
transmutation of spin orientation is compensated by that of orbit
angular momentum during the carriers moving in the system.
Although the spin current is not conserved, the total angular
momentum consisted of both spin and orbit angular momenta should
be conserved if there is no additional external force and torque.

In this paper, we inherit the Noether's idea\cite{Itzykson}, and
provide a systematical formulation of the conservation laws for
the translational and rotational motions of spin carriers based
upon the generic symmetry principles. Considering a system with
Dirac fermions coupled to an electromagnetic field, we formulated
the practical forms of spin and orbit angular currents with the
symmetries being maintained in the sense of the validity of
Noether's theorem. The nonrelativistic approximation (NRA) of
conserved currents is derived with the conservation laws being
guaranteed. Under this formalism, the torque on the angular
momentum and the force on spin carriers can be identified
rigorously. As a result we gain an insight of torque on the
transmutation of spin orientation and orbit angular momentum, and
ascertain the role of SO coupling in the conservation laws.

Consider first the relativistically covariant Lagrangian of a
closed system involved Dirac fields $\overline{\Psi }$, $\Psi $,
and electromagnetic field $A_{\mu }$. The corresponding Lagrangian
is given by $\mathcal{L}=-({\hbar c}/{2})\overline{\Psi }\gamma
_{\mu }{ \overleftrightarrow{\mathcal{D}}_{\mu }}\Psi
-mc^{2}\overline{\Psi }\Psi -({1 }/{4})\mathcal{F}_{\mu \nu
}\mathcal{F}_{\mu \nu }$\cite{Itzykson,Sakurai}, where the
four-vector notations have been used ($\mu =1,2,3,4$): $x_{\mu
}=(\mathbf{x},ict)$, $ A_{\mu }=(\mathbf{A},i\varphi )$, and
$\gamma _{\mu }=\left( \left(
\begin{array}{cc}
0 & -i\sigma _{\alpha } \\
i\sigma _{\alpha } & 0
\end{array}%
\right) ,\left(
\begin{array}{cc}
I & 0 \\
0 & -I%
\end{array}%
\right) \right) $ , $\sigma _{\alpha }$ ($\alpha =x$, $y$, $z$)
are Pauli matrices; $f{\overleftrightarrow{\mathcal{D}}_{\mu
}}g=f\mathcal{D}_{\mu }g-( \mathcal{D}_{\mu }^{\dag }f)g$,
$\mathcal{D}_{\mu }=\partial _{\mu }-i(e/\hbar c)A_{\mu }$ is a
covariant derivation operator and $\mathcal{F} _{\mu \nu
}=\partial _{\mu }A_{\nu }-\partial _{\nu }A_{\mu }$ is the
Maxwell field-strength tensor. From the Euler-Lagrangian partial
differential equation, we can obtained the Dirac equation and the
Maxwell's equation, $\gamma _{\mu }\mathcal{D}_{\mu }\Psi
+(mc/\hbar )\Psi =0$ and $\partial _{\nu }\mathcal{\ F}_{\mu \nu
}=ie\overline{\Psi } \gamma _{\mu }\Psi $, respectively.

According to Noether's theorem, any continuous one-parameter set
of invariance of Lagrangian is associated with a local conserved
current. For a closed system involved Dirac field and the
electromagnetic field, the symmetries of system yield the
conservation laws : (1) The U(1) symmetry leads to the charge
conservation. Under the local U(1) gauge transformations, $\Psi
\rightarrow \exp({-ie\Lambda /\hbar c})\Psi $, $\overline{\Psi
}\rightarrow \overline{\Psi }\exp({ie\Lambda /\hbar c})$, and
$A_{\mu }\rightarrow A_{\mu }-\partial _{\mu }\Lambda $ ,
Noether's theorem gives rise to the continuity equation $\partial
_{\mu }J_{\mu }=0$ with the charge current density $ J_{\mu
}=iec\overline{\Psi }\gamma _{\mu }\Psi $; (2) Spatial
translational isotropy begets the conservation of linear momentum
while a time-translation symmetry gives rise to the energy
conservation. The continuity equation $\partial _{\mu }\Theta
_{\mu \nu }=0$ is obtained from the invariance of Lagrangian under
the time-space translation transformations $\delta x_{\mu }=\delta
_{\mu \nu }a_{\nu }$, $\delta \Psi =0$ , $\delta \overline{\Psi
}=0$, and $\delta A_{\mu }=0$. $\Theta _{\mu \nu }=({\hbar
c}/{2})\overline{\Psi }\gamma _{\mu
}{\overleftrightarrow{\mathcal{ D}}_{\mu }}\Psi +\mathcal{F}_{\mu
\rho }\mathcal{F}_{\nu \rho }-\frac{1}{4} \mathcal{F}_{\mu'\nu'
}\mathcal{F}_{\mu' \nu' }\delta _{\mu \nu }+\partial _{\rho
}(\mathcal{F}_{\mu \rho }A_{\nu })$ is the energy-momentum current
density tensor; (3) Rotational isotropy gives birth to the
conservation of angular momentum. The Lagrangian is invariant
under the transformations $x_{\mu }^{\prime }=(\delta _{\mu \nu
}+\epsilon _{\mu \nu })x_{\nu }$, $\Psi ^{\prime }(x^{^{\prime
}})=[I-i(1/4)\epsilon _{\mu \nu }\sigma _{\mu \nu }]\Psi (x)$,
$\overline{\Psi }^{\prime }(x^{\prime })=\overline{\Psi
}(x)[I+i(1/4)\epsilon _{\mu \nu }\sigma _{\mu \nu }]$, and $A_{\mu
}^{^{\prime }}(x^{\prime })=(\delta _{\mu \nu }+\epsilon _{\mu \nu
})A_{\nu }(x)$, with $\sigma _{\mu \nu }=i(1/2)[\gamma _{\mu
,}\gamma _{\nu }]$. The total angular momentum current density
tensor $J_{\lambda ,\mu \nu }=\{\ x_{\mu }\Theta _{\lambda \nu
}+\mathcal{F}_{\lambda \mu }A_{\nu }-(\mu \leftrightarrow \nu
)\}-i({\hbar c}/{4})\overline{\Psi } \{\sigma _{\mu \nu },\gamma
_{\lambda }\}\Psi $ can be defined to satisfy the continuity
equation $\partial _{\lambda }J_{\lambda ,\mu \nu }=0$.

The NRA of Dirac equation can be carried out by separating the
Dirac wavefunction into large and small components\cite{Sakurai},
i.e., $\Psi =(\psi _{A},\psi _{B})^{T}\exp[{-i({mc^{2}+\epsilon
})t/\hbar }]$. Following the procedure of NRA given in Ref.[17],
the normalized two-component wavefunction can be written in the
form of ${\psi }=[1-(\hbar^2/8m^{2}c^{2}){\sigma }\cdot \mathbf{D
}{\sigma }\cdot \mathbf{D }]\psi _{A}$ with a notation
$\mathbf{D}=\nabla -i(e/\hbar c)\mathbf{A}$. The NRA Hamiltonian
reads\cite{Sakurai,Winkler04} ${H}=-(\hbar^2/2m)\mathbf{D
}^{2}+e\varphi +i({e\hbar^2 }/{4m^{2}c^{2}}){\sigma } \cdot \left(
\mathbf{E}\times \mathbf{D }\right) -({e\hbar }/{2mc}){\ \sigma
}\cdot \mathbf{B}-({e\hbar ^{2}}/{8m^{2}c^{2}})\nabla \cdot
\mathbf{E} -i({e\hbar^2 }/{8m^{2}c^{2}}){\sigma }\cdot \left(
\nabla \times \mathbf{E} \right) $, where $\mathbf{E}=-\nabla
\varphi $ and $\mathbf{B}=\nabla \times \mathbf{A}$ are the
electric field and the magnetic field, respectively.

Alternative to obtain the continuity equation from Hamiltonian we
engage the same procedure of NRA to the relativistic continuity
equations. The corresponding NRA of currents can be obtained from
the relativistic continuity equations by the power expansion in
$v^2/c^2$ with the same procedure as obtaining the Hamiltonian
with a SO coupling term. The symmetries of system are not broken
by this power expansion. It is instructive to demonstrate it in
the NRA of charge current $J_{\mu }=iec\overline{\Psi }\gamma
_{\mu }\Psi $, which satisfies $\partial _{\mu }J_{\mu }=0$ in the
relativistic quantum mechanics. In the nonrelativisitic limit to
the order of $ v^{2}/c^{2}$, the continuity equation becomes
$\dot{\rho}^{ \mathcal{C}}+\nabla \cdot
\mathbf{J}^{\mathcal{C}}=0$. The charge density $\rho ^{
\mathcal{C}}=e{\psi }^{\dag }{\psi }+({\hbar
^{2}e}/{8m^{2}c^{2}})\nabla ^{2}({\psi }^{\dag }{\psi })-i({\hbar
^{2}e}/{8m^{2}c^{2}})\nabla \cdot ({ \psi }^{\dag }\sigma \times
{\overleftrightarrow{\mathbf{D}}}{\psi })$ and the current density
$ \mathbf{J}^{\mathcal{C}}=-i({\hbar e}/{2m}){\psi }^{\dag }{
\overleftrightarrow{\mathbf{D}}}{\psi }+({e^{2}\hbar
}/{4m^{2}c^{2}})[ \mathbf{E}+{({2mc^{2}}/{e})}\nabla ]\times {\psi
}^{\dag }\sigma {\psi }$ have been expressed in terms of the basic
local observables. Look at the charge density first. The second
term concerns the variance of localized density of carriers. The
last term is in connection with the accumulation of localized spin
due to the spin flip. These two terms give a polarization vector
\begin{equation}
{\bf P}=-{{\hbar ^{2}e}\over{8m^{2}c^{2}}}[\nabla ({\psi }^{\dag
}{\psi })-i{ \psi }^{\dag }\sigma \times
{\overleftrightarrow{\mathbf{D}}}{\psi }].
\end{equation}
This shows that the nonuniform and the SO interaction in the
system contribute an effective electric dipole. In the charge
current the first two terms are recognized as identical to $e\rho
^{\mathcal{C}}\mathbf{v}$ with the Hamiltonian given above, where
$\bf v$ is the velocity vector. The last term is a "spin
magnetization current" $c\nabla \times \mathbf{m}$ arising from an
"intrinsic magnetic moment", where $\mathbf{m} =(e\hbar
/2mc)({\psi }^{\dag }\sigma {\psi })$. The supplement of a
magnetization term in the charge current has been established for
a long time\cite{Gurtler75}. The Dirac equation implies the
conservation law for the current $\partial _{\mu }J_{\mu
}^{\mathcal{C}}=0$ as well as the decomposition $J_{\mu
}^{\mathcal{C}}=J_{\mu ,G}^{\mathcal{C}}+\partial _{\nu }m_{\mu
\nu }$, where $J_{\mu ,G}^{\mathcal{C}}$ is Gordon current and
$m_{\mu \nu }=i(e\hbar /4mc){\overline{\Psi }}[\gamma _{\mu
},\gamma _{\nu }]\Psi $ is interpreted as the magnetic moment
density due to the carrier spin\cite{Gurtler75}.

The above analytic approach can be applied to the momentum current
tensor $\Theta_{\mu \nu }$. From $\partial _{\mu }\Theta _{\mu
\alpha}=0$ we have the continuity equation
$\mathbf{\dot{\rho}}^{\mathcal{M}}+ \mathbf{\nabla }\cdot
\mathbb{J}^{\mathcal{M}}=0$, where the momentum density is given
by
\begin{eqnarray}
\rho^{\mathcal{M}} &=&-i{{\hbar}\over{2}}(1+{{\hbar
^{2}}\over{8m^{2}c^{2}}}\nabla ^{2}){\psi }^{\dag
}\overleftrightarrow{\bf D}{\psi
}+\rho^{\mathcal{M}}_{SO}\notag\\
&&+\rho^{\mathcal{M}}_G-{{\hbar ^{2}e}\over{8m^{2}c^{3}}}(\nabla
\rho^{\cal C})\times {\bf B}+{1\over c}\mathbf{E}\times \mathbf{B}
\end{eqnarray}
with $\rho^{\mathcal{M}}_{SO}=-i({\hbar^2 e}/{8m^{2}c^{3}}){\psi
}^{\dag }(\sigma {\bf B}\cdot\overleftrightarrow{\bf
D}-\sigma\cdot {\bf B}\overleftrightarrow{\bf D}){\psi }$ and
$\rho^{\mathcal{M}}_G=({\hbar }/{16m^{2}c^{2}})\nabla\cdot( {\psi
}^{\dag }\sigma \times {\overleftrightarrow{{\cal{\bf D}}{\cal
{\bf D}}}} {\psi })-({\hbar ^{2}e}/{8m^{2}c^{3}})\nabla
\times(\psi^\dag\psi{\bf B})+(1/ c)\mathbf{\nabla }\cdot
(\mathbf{EA})$. $\mathbb{J}^{\mathcal{M}}$ is the nonrelativistic
limit of momentum current which takes a form as
\begin{eqnarray}
\mathbb{J}^{\mathcal{M}} &=&-\frac{\hbar ^{2}}{4m}{\psi }^{\dag }
\overleftrightarrow{\mathbf{DD}}{\psi }-i\frac{\hbar ^{2}e}{
8m^{2}c^{2}}[\mathbf{E}\times ({\psi }^{\dag }\mathbf{\sigma }
\overleftrightarrow{\mathbf{D}}{\psi })
\notag\\
&& +({\psi }^{\dag }\mathbf{\sigma }\times
\overleftrightarrow{\mathbf{D}} {\psi })\mathbf{E -\mathbf{\nabla
}({\psi }^{\dag } {\psi })\mathbf{E}+(\mathbf{\nabla E }){\psi
}^{\dag }{\psi }]} \notag\\
&&-(\mathbf{EE}+\mathbf{BB}) +\frac{1}{2}(E^2+B^2-\frac{\hbar
e}{mc}\mathbf{B}\cdot {\psi } ^{\dag }\mathbf{\sigma }{\psi
})\mathbb{I} \notag\\
&& +\frac{\hbar e}{2mc}\mathbf{B}{\psi }^{\dag }\mathbf{\sigma }
{\psi }+\mathbf{\nabla }\times (\mathbf{BA}-i\frac{\hbar
^{2}}{4m}\mathbf {\psi }^{\dag } \mathbf{\sigma
}\overleftrightarrow{\mathbf{D}}{\psi })
\end{eqnarray}
with the notation ${\psi }^{\dag }{\overleftrightarrow{D_{\beta
}D_{\alpha }}}{\ \psi }={\psi }^{\dag }D_{\beta }D_{\alpha }{\psi
}+(D_{\beta }^{\ast }D_{\alpha }^{\ast }{\psi }^{\dag }){ \psi
}-(D_{\alpha }^{\ast }{\psi }^{\dag })(D_{\beta }{\psi
})-(D_{\beta }^{\ast }{\psi }^{\dag })(D_{\alpha }{\psi })$.

The NRA of conservation of total angular momentum current tensor
$J_{\lambda ,\mu \nu }$ can be obtained in the same way. The local
angular momentum density is found as
\begin{equation}
\mathbf{\rho }^{\mathcal{A}}={\widetilde{\bf x}}\times
\mathbf{\rho }^{\mathcal{M}}+\rho^{S}+{1\over c}\mathbf{E}\times
\mathbf{A}+{{\hbar ^{2}e}\over {4m^{2}c^{3}}}\rho^{\cal C}{\bf B},
\end{equation}
where ${\widetilde{\bf x}}={\bf x}-{\bf x}_0$ with a constant
vector ${\bf x}_0$. The first term is the conventional angular
momentum with the momentum density, while the last two terms in
Eq. (4) come from electromagnetic field. The second term is the
intrinsic angular momentum density, associated with the local
spin, given by $\rho^{\cal S} =({\hbar }/{2})[1-({\hbar
^{2}}/{8m^{2}c^{2}})\nabla ^{2}]{\psi }^{\dag }\sigma {\psi
}+\rho^{\cal S}_{SO}+\rho^{\cal S}_G$, where $\rho^{\cal S}_{SO}
=({\hbar ^{3}}/{8m^{2}c^{2}})[2{\psi }^{\dag }\sigma{\bf D}^2{\psi
}+({\cal {\bf D}}^\dag{\psi }^{\dag })\cdot {\sigma }({\cal {\bf
D}}{\psi }) +({\cal {\bf D}}^\dag{\psi }^{\dag }){\sigma
}\cdot({\cal {\bf D}}{\psi })]$ is the contribution of SO coupling
induced by the transmutation and $\rho^{\cal S}_G$ relates to the
surface terms $\nabla \cdot ( {\psi } ^{\dag }\sigma
\overleftrightarrow{\cal{\bf D}}{ \psi })$ and $\nabla\times (
{\psi } ^{\dag }\overleftrightarrow{\cal{\bf D}}{\psi })$. The
local angular momentum density satisfies the continuity equation
$\dot{\mathbf{\rho }}^{\mathcal{A}}+\mathbf{\nabla }\cdot
\mathbb{J} ^{\mathcal{A}}=0$ with the total angular momentum
current density tensor
\begin{eqnarray}
\mathbb{J}^{\mathcal{A}} &=&\widetilde{\bf x}\times
[\mathbb{J}^{\mathcal{M}}+i\frac{\hbar ^{2}}{4m} \mathbf{\nabla
}\times ({\psi }^{\dag }\mathbf{\sigma }
\overleftrightarrow{\mathbf{D}}{\psi })] -i\frac{ \hbar
^{2}}{4m}{\psi }^{\dag }\overleftrightarrow{\mathbf{D}}
\mathbf{\sigma }{\psi }\notag\\
&& +i\frac{\hbar ^{2}}{4m}\mathbf{\nabla }\times [{\psi }^{\dag }
\mathbf{\sigma }\overleftrightarrow{\mathbf{D}}{\psi }\times
\widetilde{\bf x}]+(\mathbf{B}\cdot
\mathbf{A)}\mathbb{I}-\mathbf{AB},
\end{eqnarray}
where $\mathbb{I}$ is unit tensor. Notice that the survival of the
term adhered the constant vector $\mathbf{x}_0$ in a conserved
angular momentum of system is not a trivial matter. The continuity
equation of angular momentum is satisfied with the symmetry of the
invariance under spatial translation being guaranteed. Let us
emphasize that the present form of total angular current is
independent of the choice of coordinate origin in the system.

So far we have got the NRA expression of the conserved angular
momentum current for a closed system including the fermion and the
electromagnetic field. In general, the NRA conserved currents
consist of three parts: orbital, spin, and electromagnetic. We
next give a decision in favour of the angular momentum current of
fermionic system by taking the electromagnetic field as an
external field. Because the conservation laws are equations of
motion for the observables, the force and the torque on the spin
carriers, coupled to the external field, can be determined by the
continuity equations. It is noted $\nabla \cdot \lbrack
(\mathbf{EE}+\mathbf{BB})-({1}/{2})(E^2+B^2)\mathbb{I}+\mathbf{
\nabla }\times (\mathbf{BA})]=\mathbf{E}\rho
^{C}+({1}/{c})(\mathbf{J}^{ \mathcal{C}}\times \mathbf{B})$, which
corresponds to Lorentz force, in the continuity equation of
momentum flux when the Maxwell equation is applied to the
time-independent situation. While the terms, in Eq.(3), $-({\hbar
^{2}e}/{8m^{2}c^{2}})[i(\psi ^{\dag }\mathbf{\sigma } \times
\overrightarrow{\mathbf{D}}\psi )\mathbf{E}-\mathbf{\nabla }(\psi
^{\dag }\psi ) \mathbf{E}+(\mathbf{\nabla E})\psi ^{\dag }\psi]
+({\hbar e}/{2mc})[ \mathbf{B}\psi ^{\dag }\mathbf{\sigma }\psi
-(\mathbf{B} \cdot \psi ^{\dag }\mathbf{\sigma }\psi )\mathbb{I}]$
can be regarded as the source of the force associated with the
carrier's spin\cite{Ryu96,Barone73}. The continuity equation of
momentum current becomes $\mathbf{\dot{\rho}}{^{
\mathcal{M}}}^{\prime }+\mathbf{\nabla }\cdot
\mathbb{J}{^{\mathcal{M}}} ^{\prime }=\mathbf{F}$, where the
primes are meant to indicate the fermionic partition of density
and current. The momentum density vector and the momentum current
density tensor are given by
$\mathbf{{\rho}}{^{\mathcal{M}}}^{\prime }=-i({\hbar }/{2})\psi
^{\dag }\overleftrightarrow{\mathbf{D}}\psi $ and $
\mathbb{J}^{\mathcal{M}\prime }=-({\hbar ^{2}}/{4m})\psi ^{\dag }
\overleftrightarrow{\mathbf{DD}}\psi -i({\hbar
^{2}e}/{8m^{2}c^{2}}) \mathbf{E}\times (\psi ^{\dag
}\mathbf{\sigma }\overrightarrow{\mathbf{D}} \psi )-i({\hbar
^{2}}/{4m})\mathbf{\nabla }\times (\psi ^{\dag }\mathbf{ \sigma
}\overleftrightarrow{\mathbf{D}}\psi )$, respectively. The total
force is found

\begin{eqnarray}
\mathbf{F} &=&e(\mathbf{E}+\frac{\hbar
^{2}}{8m^{2}c^{2}}\mathbf{\nabla }^{2}\mathbf{E}){\psi}^{\dag
}{\psi} +[{\bf J}^{\cal C}-c\nabla\times{\bf m}]\times {\bf B}
\notag\\
&&+\frac{\hbar e}{2mc}(\mathbf{\nabla B})\cdot \rho^{\cal
S}+i\frac{\hbar ^{2}e}{8m^{2}c^{2}}(\mathbf{\nabla E})\cdot
{\psi}^{\dag }\mathbf{\sigma \times
}\overleftrightarrow{\mathbf{D}}{ \psi }.
\end{eqnarray}
The first three terms are the Lorentz force and  the
"Stern-Gerlach force" which relates to a magnetic field gradient
$\mathbf{\nabla B}$. The forth term is new which has not been
discussed previously. It relates to the fluctuation of localized
spin current and produces a torque on the angular momentum. The
continuity equation of angular momentum has the general form
$\mathbf{\dot{\rho}}^{\mathcal{A}\prime}+\mathbf{\nabla }\cdot
\mathbb{J}^{\mathcal{A}\prime }=\mathbf{M}$, where
$\mathbb{J}^{\mathcal{A}\prime }$ is the total angular momentum
current of electron which can be written in the form of a sum of
three terms $ \mathbb{J}^{\mathcal{A}\prime
}=\mathbb{J}^{\mathcal{O}}+\mathbb{J}^{ \mathcal{S}}+i({\hbar
^{2}}/{4m})\nabla\times({\psi }^{\dag }\sigma
{\overleftrightarrow{\bf D}}{\psi }\times \widetilde{\bf x}) $.
The orbit angular momentum current is given by
\begin{equation}
\mathbb{J}^{\mathcal{O}}=\widetilde{\bf x}\times
[\mathbb{J}^{\mathcal{M}\prime }-i{{\hbar ^{2}}\over
{4m}}\mathbf{\nabla }\times (\psi ^{\dag }\mathbf{ \sigma
}\overleftrightarrow{\mathbf{D}}\psi )]
\end{equation}
and the spin current is defined by
\begin{equation}
J_{\beta \alpha }^{\mathcal{S}}=-i\frac{\hbar ^{2}}{4m}\psi ^{\dag
}\sigma _{\alpha }{\overleftrightarrow{D_{\beta }}}\psi -\epsilon
_{\beta \alpha \gamma }\frac{\hbar ^{2}e}{8m^{2}c^{2}}E_{\gamma
}\psi ^{\dag }\psi.
\end{equation}
The term on the right hand side of continuity equation is the
torque $\mathbf{M}$ which relates to the total angular momentum,
not spin partition itself. We find
\begin{equation}
\mathbf{M}=\widetilde{\mathbf{x}}\times \mathbf{F}+\frac{
e}{mc}{\bf \rho}^{\cal S}\times \mathbf{B}+i\frac{\hbar
^{2}e}{8m^{2}c^{2}}({ \psi }^{\dag }\mathbf{\sigma }\times
{\overleftrightarrow{\mathbf{D}}}{\psi } )\times \mathbf{E}.
\end{equation}
There are several origins in the torque $\mathbf{M}$. Besides the
conventional torque caused by the force ${\bf F}$, the second term
is the torque in the fundamental equation of spin resonance theory
and the third term is a torque induced by the spin current.
Because we are ultimately interested in relating the induced
torque to the effect of SO coupling, a fluctuation in the spin
current ${\Delta \mathbf{J}^{\mathcal{S}}}=i({\hbar
^{2}}/{4m})\psi ^{\dag }(\sigma \times {\overleftrightarrow{\bf D}
)}\psi $ is introduced or, equivalently
\begin{eqnarray}
\Delta J_{\alpha }^{\mathcal{S}}=\epsilon _{\alpha \beta \gamma
}J_{\beta \gamma }^{\mathcal{S}}+{\frac{\hbar
^{2}e}{4m^{2}c^{2}}}E_{\alpha }\psi ^{\dag }\psi.
\end{eqnarray}
By using Eq. (8) the relation between $\Delta J_{\alpha
}^{\mathcal{S}}$ and the spin current can be put in the form
$\epsilon _{\alpha \beta \gamma }\Delta J_{\gamma
}^{\mathcal{S}}=J_{\alpha \beta }^{\mathcal{S}}-J_{\beta \alpha
}^{\mathcal{S}}$. This vector has its own
significance\cite{Rashba03}. It contributes a force if there is a
spatial change of electric field and a torque on the spin when the
carriers moves in an electric field. From Eqs. (6) and (9) an
force $\Delta \mathbf{F}$ and a torque $\Delta \mathbf{M}$ induced
by the SO coupling may be obtained, which can further be separated
into the contributions from the accumulation of spin and the
fluctuation of spin current, in a compact form

\begin{eqnarray}
\left(
\begin{array}{c}
\Delta \mathbf{F} \\
\Delta\mathbf{M}
\end{array}
\right) =\frac{e}{2mc^{2}}\left(
\begin{array}{cc}
2c(\nabla \mathbf{B})\cdot & (\nabla \mathbf{E})\cdot \\
-2c\mathbf{B}\times& -\mathbf{E}\times
\end{array}%
\right) \left(
\begin{array}{c}
\mathbf{\rho }^{\mathcal{S}} \\
\Delta \mathbf{J}^{\mathcal{S}}
\end{array}%
\right).
\end{eqnarray}
It is noted that $\Delta \mathbf{J}^{\mathcal{S}}$ is just the
"electric dipole" $<{\hat {\bf l}}>= -<({\hat{\mathbf{\mu
}}}/{2c}){\times }({-i\hbar {\hat{\mathbf{D}}}}/{m})>$  with the
intrinsic magnetic dipole moment ${\hat {\mathbf{\mu }}}=({\hbar
e}/{2mc})\mathbf{\sigma }$. Physically, it is thinking to be a
spin vector rotates around an axis with a velocity $-i\hbar
(\mathbf{D}/{m})$ to give rise a magnetic current circuit,
equivalently, an "electric dipole". The torque on the "electric
dipole" is $<{\hat{\bf l}}>\times {\bf E}$ as expected.

Next, we apply the formulism to a two-dimensional electron gas
(2DEG) with the Rashba SO interaction. This can be achieved by
taking $ {\bf E}=(0,0,E_{z})$ and ${\bf A}=0$ in our derivations.
In the follow discussion the indices $k$ and $l$ are specialized
to 2D indices while $\alpha$ still is indicated in 3D space. The
Rashba Hamiltonian is then given by
$H_{R}=({1}/{2m})\mathbf{p}^{2}-(\lambda /\hbar )(p_{x}\sigma
_{y}-p_{y}\sigma _{x})$ with $\lambda =({\hbar ^{2}e}/{\
4m^{2}c^{2}})E_{z}$. The momentum current in Eq. (3) reduces to
${J_{kl }^{\mathcal{M}}}^{\prime } =-({\hbar ^{2}}/{4m}){\psi }
^{\dag }{\overleftrightarrow{\nabla_{k }\nabla _{l }}}{\psi }
+i\epsilon _{kj z}({\lambda }/{2})[{\psi }^{\dag }\sigma
_{j}{\overleftrightarrow{\nabla}_{l }}{\psi } -({\hbar
^{2}}/{2m\lambda})\partial _{j }({\psi } ^{\dag }\sigma
_{z}{\overleftrightarrow{\nabla }_{l }}{\psi })]$. It is easy to
prove that $\nabla _{k }{J_{kl }^{\mathcal{M}}} ^{\prime }=0$ and
$F_{\beta }=0$ for the stationary states. The total angular
momentum current is given by
\begin{equation}
{J}_{k \alpha }^{\mathcal{A}\prime }={J}_{k\alpha
}^{\mathcal{O}}+{J}_{ k\alpha k}^{ \mathcal{S}}+i{{\hbar
^{2}}\over {4m}}\nabla\times[({\psi }^{\dag }\sigma
{\overleftrightarrow{\bf D}}{\psi })\times \widetilde{\bf x}]_{k
\alpha}.
\end{equation}
It is shown that the orbit part of angular momentum current
density, ${J}_{k\alpha}^{\mathcal{O}}=-({\hbar
^{2}}/{4m})\epsilon_{\alpha jl}x_j({\psi } ^{\dag
}{\overleftrightarrow{\nabla_{l }\nabla _{k }}}{\psi })-i({\lambda
}/{2})\delta_{\alpha z}{\widetilde{\bf x}}\cdot({\psi }^{\dag
}\sigma {\overleftrightarrow{\nabla}_{k }}{\psi })$, is
nonvanishing only for $\alpha =z$. The corresponding spin current
is given by $J_{k \alpha }^{\cal S}=-i({\hbar ^{2}}/{4m}){\psi
}^{\dag }\sigma _{\alpha }{ \overleftrightarrow{\nabla }_{k
}}{\psi }-\epsilon _{k \alpha z} ({\lambda }/{2})({\psi }^{\dag
}{\psi })$. It is shown that the divergence of spin current, in
general, satisfies the equation
$\dot{\mathbf{\rho }}^{\mathcal{S}}+\mathbf{\nabla }\cdot \mathbb{J}^{%
\mathcal{S}}=\mathbf{T}$ with the torque $T_{x(y)}=i({\lambda
/2})({\psi }^{\dag }\sigma
_{z}\overleftrightarrow{\nabla}_{x(y)}{\psi })$ and
$T_{z}=-i({\lambda /2} )({\psi }^{\dag }\sigma \cdot
\overleftrightarrow{\nabla} { \psi })$. A straightforward
calculation shows $\nabla_{k }J_{k z}^{\cal O}=-T_{z}$. This leads
to $\nabla_{k }J_{k z}^{A\prime }=0$, while $\nabla_{k
}J_{kl}^{A\prime }=T_l$, i.e. the component of total angular
momentum perpendicular to the x-y plane is conserved, which shows
clearly how the transmutation between the spin and the orbit
angular momenta occurs in the system with a SO coupling, while the
in-plan components of spin current are not conserved by requiring
a torque on it. It is worth to emphasize that ${\bf F}$ is
nonvanishing in the presence of an in-plan electric field. The
force gives rise to an extra conventional torque. In general, it
can be shown a relation between the torques ${\bf T}$ and ${\Delta
{\bf M}}$, acting on the spin itself and on the total angular
momentum, respectively,
\begin{equation}
\mathbf{\Delta M}=\mathbf{T}+i\frac{\hbar ^{2}e}{8m^{2}c^{2}}\psi
^{\dag }( \mathbf{\sigma \times E})\times
\overleftrightarrow{\mathbf{D}}\psi.
\end{equation}

We note that there is an essential difference between the force
$\bf F$ defined from the continuity equation of momentum current
and the transverse force $\bf f$ on the spin for driving spin
current. The latter is the covariant force on a spin defined from
the equation of motion
\begin{equation}
m\dot{\mathbf{v}}=-m\mathbf{\nabla }\cdot \mathbb{J}^{ v}+{
\mathbf{f}},
\end{equation}
where $\mathbf{v}$ is the velocity of a wave packet of
spin-polarized electron. In a Rashba 2DEG it is found
${v}_j=-({\hbar^2}/{2m})\psi ^{\dag }\overleftrightarrow{D
}_j\psi-({2\lambda}/{\hbar^2})\epsilon_{zjk}\rho^{\cal
S}_k+(1/m)\epsilon_{jkl}\nabla_k\rho^{\cal S}_l$. ${\mathbb{
J}}^{v}$ is the corresponding tensor of velocity flux given by
${J}_{jk}^{v}=(1/m){J}_{jk}^{\mathcal{M}\prime
}+(2\lambda/\hbar^2)\epsilon_{zlk}{J}_{jl}^{\mathcal{S}}$. The
second term in the velocity flux can be understood as a drift
velocity because of bringing the electric field to bear on the
moving spin. The covariant force on a spin is found as
\begin{equation}
{f}_j= -\frac{2m\lambda}{\hbar^{2}}\epsilon_{zjk}T_k
+\epsilon_{jk\alpha}{\nabla }_k{T}_\alpha.
\end{equation}
Replacing $T_k=-(2m\lambda/\hbar^2){J}_{kz}^{\cal S}$ into the
second term of ${f}_j$, we find
${f}_j=(4m^2\lambda^2/\hbar^4)\epsilon_{zjk}J^{\cal S}_{kz}$.
Because of the opposite signs for the contrary components of
spin-polarized spin current, the different spins experience the
transverse forces in the opposite
directions\cite{Nikolic05,Shen05}.

The main results of this paper may now be summarized as follows.
The conservation law of NRA conserved angular momentum based on
the Noether's theorem has been formulated systematically and
rigorously. It is identified with that the torques are generated
on a moving spin through the "electric dipole" when the
electromagnetic field is regarded as an external field. As an
example of the application, we consider a Rashba 2DEG and show
that a torque will develop on the spin. The transmutation between
the spin and the orbit angular momenta has been demonstrated
explicitly. Furthermore, the covariant force on the spin in the
presence of SO coupling is clarified.

\begin{acknowledgments}
This work is supported by NNSFC grants 10274069 and 10474002.
\end{acknowledgments}
\noindent{$^{a)}$ Electronic mail: mazs@phy.pku.edu.cn}
\bibliographystyle{plain}

\end{document}